\documentclass{article}

\usepackage[OT2, T1]{fontenc}
\usepackage[english]{babel}
\usepackage{graphicx}
\usepackage{authblk}
\usepackage[lmargin=3cm,rmargin=3cm,tmargin=2cm,bmargin=3cm,marginpar=2cm ,reversemp]{geometry}
\usepackage{amsmath,amssymb,bm,color,mathrsfs,wasysym}
\usepackage{hyperref}

\begin{document}

\title{Low temperature nucleation rate calculations using the N-Fold way}

\date{}

\author[1]{Federico Ettori
\thanks{federico.ettori@polimi.it}}
\author[2,3]{Dipanjan Mandal
\thanks{dipanjan.mandal@univaq.it}}
\author[2]{David Quigley
\thanks{d.quigley@warwick.ac.uk}}

\affil[1]{\small Department of Physics, Politecnico di Milano, Piazza Leonardo da Vinci 32, 20133 Milan, Italy}
\affil[2]{\small Department of Physics, University of Warwick, Coventry, CV4 7AL, United Kingdom}
\affil[3]{\small Department of Physical and Chemical Sciences, University of L’Aquila, Via Vetoio, 67100 L’Aquila, Italy}

\maketitle

\begin{abstract}
    \noindent We present a numerical study to determine nucleation rates for magnetisation reversal within the Ising model (lattice gas model) in the low-temperature regime, a domain less explored in previous research. To achieve this, we implemented the N-Fold way algorithm, a well-established method for low-temperature simulations, alongside a novel, highly efficient cluster identification algorithm.  Our method can access nucleation rates up to 50 orders of magnitude lower than previously reported results. We examine three cases: homogeneous pure system, system with static impurities, and system with mobile impurities, where impurities are defined as sites with zero interactions with neighbouring spins (spin value of impurities is set to 0). Classical nucleation theory holds across the entire temperature range studied in the paper, for both the homogeneous system and the static impurity case. However, in the case of mobile impurities, the umbrella sampling technique seems ineffective at low mobility values. These findings provide valuable insights into nucleation phenomena at low temperatures, contributing to theoretical and experimental understanding.
\end{abstract}

\section{Introduction}
Nucleation is an out-of-equilibrium process describing the initial stages of the formation of a new thermodynamic phase from a metastable parent phase. For solutions and mixtures, the process involves structural reorganisation of the constituent components. Such phenomenon takes place in various real systems ranging from biomineralisation~\cite{Jin2018,Weiner2011} to magnetic reversal processes~\cite{Vogel2006}, and its regulation could be useful for many applications, such as pharmaceutical manufacture~\cite{Lee2011}, synthesis of nanoelectronics devices~\cite{Sosso2013,Yarema2018} and climate modelling~\cite{Zhang2012,Khvorostyanov2000}.

Classical Nucleation Theory~\cite{Dimo} (CNT) is a well-established theory to analyse nucleation phenomena and provide quantitative estimates and predictions that can be used as support for experimental results. Within this framework, one of the most investigated parameters is the nucleation rate~\cite{Blow2021}, defined as the number of nucleation events per unit time and volume. This rate is typically estimated with Becker-D\"oring-Zeldovich (BDZ) theory~\cite{Becker1935,Zeldovich1943}. Despite strong qualitative agreements, theory and experiments often exhibit quantitative discrepancies~\cite{Zimmermann2015}, arising from difficulties of the theory to incorporate essential microscopic behaviour characterising the nucleation process, such as particle dynamics, diffuse interfaces or the presence of impurities in the system.

Computer simulations represent an important tool in helping to test and improve theoretical predictions, as the controlled environment and the possibility of following closely the formation of the equilibrium phase are of significant value in challenging the basic CNT description. Molecular dynamics~\cite{Anwar2011,Desgranges2009} is a common choice particularly suitable for modelling the microscopic characteristics of a reduced number of particles. CNT approximations are applied with difficulty in this case, where it is often unclear what can be regarded as the volume or the surface of a highly irregular cluster. A different approach is given by Monte Carlo simulations employed in 3D and 2D lattice gas models~\cite{Stauffer1982,Binder2016}. These are particularly suitable to model systems with quenched randomness and can capture both heterogeneous and homogeneous nucleation processes~\cite{Sear2006,Scheifele2013,Mandal2022,Ettori2023}.  

Of extreme importance for many nucleation processes is the variation of nucleation rate with temperature. By changing temperature, it is possible to observe variations of several orders of magnitude in nucleation rate estimates. In the low-temperature regime, thermal fluctuations are strongly reduced, and nucleation is regarded as a rare event. For this reason, we expect the microscopic characteristics of the system, such as particle mobility and geometric frustration~\cite{Mengyuan2023}, to play a major role in accelerating or slowing the transition from the metastable to the stable state. This was found by Mandal and Quigley~\cite{Mandal2021}, where nucleation at low temperatures was enhanced by introducing a small fraction of neutral dynamical impurities. Another case study at low temperatures regards the observation of an anomaly in the silicon glass nucleation rate estimation~\cite{Weinberg1989,Jurn2016}, even if it is claimed to be an artefact related to limited heating time~\cite{Xia2021}. Still, the low-temperature regime hides many interesting behaviours and needs further investigation.

Impurities (or additives) are usually observed in real systems and affect the nucleation process in many complex ways, as in the case of calcium carbonate~\cite{Gebauer2009}. Impurities can be employed to hinder the nucleation of supersaturated solutions~\cite{Han2019,Xu2024}, to accelerate it~\cite{Poon2015}, to change the crystallographic structure or the growing mechanism of the stable phase~\cite{Vorontsov2018}.

In this work, we propose an algorithm for the two-dimensional Ising (lattice-gas) model suitable for low-temperature simulations with static or mobile impurities, based on the N-Fold algorithm~\cite{Bortz1975} (also known as BKL algorithm). We test its performance and compare it with the standard Metropolis algorithm. We find a consistent increase in efficiency with a factor of 50 speed up at the lowest simulated temperatures. With this algorithm, we investigate the nucleation rate in the low-temperature regime ($0.7<T<1.6$), reaching values of $50$ orders of magnitudes lower than previous estimates available in the literature~\cite{Ryu2010,Mandal2021}. Within this regime, we report good agreement between CNT and simulations, for the homogeneous, static and fast-moving impurity cases. We finally propose a new and efficient cluster identification algorithm, optimised for nucleation rate calculation.

The outline of the work is the following. In section~\ref{sec:Model} we describe the model and the algorithms used. We define the impurities employed to mimic the real behaviour of systems subjected to impurity-moderated nucleation. In section~\ref{sec:CNT} we briefly review the main results of CNT and in section~\ref{sec:RESM} we discuss the two main techniques employed in this study to investigate rare events, namely Umbrella Sampling (US) and Forward Flux Sampling (FFS). In section~\ref{sec:Results} we present the main results of this work, starting with the time efficiency analysis between the proposed algorithm (N-Fold way) and the standard algorithm (Metropolis) according to the literature. We discuss the nucleation free energy function derived from simulations and compare it with the functional form predicted by CNT. We then present estimates for the nucleation rate and we show good agreement between both US and FFS procedures. Finally, we make concluding comments in section~\ref{sec:Conclusion}.

\section{Model and Algorithm}\label{sec:Model}
To model nucleation from a metastable phase we consider the two-dimensional Ising model on a square $L\times L$ lattice of with periodic boundary conditions in the presence of impurities. For all the presented simulations, we fix $L=100$. Each site on the square lattice is associated with a variable $s_i$, where $i$ is the site index. The total energy of the system follows from the Hamiltonian
\begin{equation}
    \mathcal{H} = -J\sum_{\langle i,j\rangle} s_is_j - h\sum_i s_i,
\end{equation}
in which $\langle i,j\rangle$ represents the nearest neighbours and $h$ is the applied external field which breaks $s_i\leftrightarrow -s_i$ symmetry. We fix the interaction strength between neighbouring sites $J=1$ throughout the paper. We set $h$ to a small positive value ($h=0.05$ unless otherwise stated, equivalent to a chemical potential difference of $\Delta \mu= 2$h$ =0.1$) such that the dominating $+1$ phase is the stable equilibrium phase and the dominating $-1$ phase is the metastable phase. The variable $s_i$ can take three different values which are $s_i=+1, -1, 0$. The model can have two different interpretations. The first is the magnetic interpretation in which $s_i=+1, -1, 0$ are respectively up, down and zero spins at site $i$. Up and down spins can flip in the presence of an external magnetic field $h$ but non-interacting neutral spins do not respond to the external field and act as impurity sites. The second interpretation of the model describes solute precipitation in which $s_i=+1$ represents a solute particle, $s_i=-1$ represents a solvent particle and $s_i=0$ represents an impurity particle at site $i$. The chemical potential difference between the nucleating phase (dominated by +1 solutes) and the solution phase (dominated by -1 solvents) in the presence of impurities can be written as $\Delta \mu \approx2h$. This approximation is valid at low temperatures and low impurity density where the solvent is sufficiently strong that there are only solute monomers present in the metastable solution phase. We will use this solute precipitation interpretation throughout the paper.

We investigate the nucleation rate starting from the metastable solution phase to the nucleating phase at low temperatures for three different scenarios: no impurities, static and dynamic impurities. In the static case, impurities are initialised with random positions on the lattice that remain fixed throughout the simulations. In the dynamic case, impurities are free to move within the lattice through nearest neighbour swap moves known as Kawasaki dynamics~\cite{Kawasaki1966}. We tune the relative movement of solute and solvent with impurities through the impurity mobility parameter $\alpha \in [0,1]$. This regulates the rate at which impurity swaps are attempted compared to solute-solvent transmutation moves (spin flips). In the following, we will refer to Impurity Swap moves (Kawasaki dynamics) with IS and Particle Update moves (transmutation) with PU. 

Various interesting roles, e.g., surfactant, inert spectator, heterogeneous nucleating sites and bulk stabiliser can be played by the impurities that influence the nucleation properties, depending on how these interact with solute and solvent~\cite{Mandal2024}. At low temperatures, impurities act as a surfactant occupying the boundary positions of a growing cluster. This helps decrease the interfacial tension and boost the nucleation rate as shown in simulations by Mandal and Quigley~\cite{Mandal2021}. However, performing rare-event simulations using the conventional Metropolis algorithm in this regime is not efficient as most attempted moves are rejected. The algorithm of choice at low temperatures is the N-Fold way~\cite{Bortz1975}, specifically adapted to overcome this problem. We also compare the results with the standard Metropolis algorithm (at the high-temperature regime) for validation and to ensure the generality of our results.

For clarity, we briefly recall the general idea behind the Metropolis algorithm. Given a specific microstate $i$, the standard Metropolis algorithm randomly generates a new microstate $j$ (a proposed next step along a Markov chain) and it accepts or rejects this with probability $P(i \to j) = p(\Delta E)$ where $\Delta E = E_{j} - E_{i}$ represents the energy difference between the new proposed state in the Markov chain and the current state. The probability $p(\Delta E)$ is usually expressed in terms of the transition rate $w(\Delta E)$ multiplied by the time unity.  The transition rates are crucial for determining the system's dynamics. Many forms can be chosen as long as the detailed balance condition is satisfied. The most widely used are the Metropolis-Hastings transition rates
\begin{equation}
    w_{MH}(\Delta E) = 
    \begin{cases}
        \frac{1}{\tau} & \text{for } \Delta E\leq 0\\
        \frac{1}{\tau}\exp{(-\beta \Delta E)} & \text{otherwise} \\
    \end{cases}
\end{equation}
even though the Glauber transition rates~\cite{Glauber1963,Stanley} are commonly used for out of equilibrium processes
\begin{equation}
    w_{G}(\Delta E) = \frac{1}{2\tau}\left(1-\tanh{\left(\beta \frac{\Delta E}{2}\right)}\right)
    \label{eq:Glauber}
\end{equation}
Here $\tau$ represents the system time constant which is taken to be equal to $1$. This $\tau$ can be considered the time incremented at each move attempt. In this paper, we will rely on the Glauber transition rates unless otherwise specified, and hence expect our N-Fold way simulations to be kinetically consistent with Glauber dynamics, rather than Metropolis-Hastings dynamics. We also define the fundamental time unit, Monte Carlo Single Step (MCSS), as the time taken for $N$ attempted moves, out of which on average $(1-\alpha)N$ moves are particle update attempts.

In the case of static impurities, PU is the only elementary move type through which the system evolves.
For mobile impurities, we consider the implementation outlined by Mandal and Quigley~\cite{Mandal2021}, in which the new state is proposed by considering either PU or IS move depending on the value of $\alpha$. It is important to note that several modifications have been introduced into the algorithm for the present study. Firstly, time is updated with each attempted move (both PU and IS), rather than exclusively for PU moves as in reference~\cite{Mandal2021}. Secondly, the swap between two impurities is treated as a discrete event, which implies a time update even though the system configuration does not change. This modification of the algorithm ensures that the number of possible IS moves remains fixed at $4L^2f$ (with $f$ the fraction of impurities in the system) throughout the simulation. This is an important requirement for directly comparing the Metropolis and the N-Fold way algorithms since the latter requires a-priori knowledge of the total number of IS and PU moves available in the lattice at every simulation step.

The N-Fold way algorithm is a rejection-free algorithm developed by Bortz and co-authors~\cite{Bortz1975} in 1975. The advantage of this algorithm lies in its rejection-free nature. The Markov chain of states is built in a way so that the newly proposed state is accepted with probability $P(i \to j)=1$. This algorithm is particularly suitable for low temperatures at which the Metropolis algorithm rejects most of the attempted moves due to low acceptance probability. 

We briefly recall the main idea behind the N-Fold way algorithm. For coherence with the original paper, we momentarily switch to magnetic notation. To generate the next state in the Markov chain, the algorithm considers all possible energy changes that a trial move could generate. For an Ising model with no impurities, a spin flip (PU) move at site $i$ produces an energy variation equal to $\Delta E(s_i) = 2s_i \sum_{\langle i,j \rangle} s_j + 2hs_i$. Since $\Delta E(s_i)$ depends only on the relative orientation of $s_i$ with respect to the external field and the state of the four neighbouring sites, it can only take ten possible values. At the beginning of the simulation, each site $s_i$ is classified into one of the ten classes based on $\Delta E(s_i)$. At each iteration, one of the ten classes is selected with probability proportional to the transition rates $w(\Delta E_k)$ (where $k$ stands for the energy variation related to the moves of the $k$-th class) and a site at which to make the move is extracted from the chosen class with uniform probability. The selected move is applied, and the class structure is updated accordingly. After each particle update move, the affected site and its neighbours are reclassified to ensure the correct assignment to the appropriate energy class. The time evolution in the N-Fold way cannot be directly measured in terms of MCSS as, unlike the Metropolis algorithm, all the proposed moves are always accepted. To ensure a consistent dynamical representation, a continuous-time approach is employed, where an appropriate expression is used to account for the equivalent (random) number of rejected moves that would have occurred in a Metropolis implementation. In an N-Fold way algorithm, the time increment $\Delta t$ can be written as
\begin{equation}
    \Delta t = -\frac{\ln(r)}{\sum_i w_i},
    \label{eq:dt_easy}
\end{equation}
where $w_i$ the rate of $i$-th possible move and $r$ is a random number uniformly distributed in the interval $0$ to $1$.

For static impurities, we follow the implementation outlined by Ettori et al.~\cite{Ettori2023}. Essentially, at the beginning of the simulation, all impurities are grouped into a class with zero transition probability to exclude impurities from transmutation moves. Additionally, during the update of the neighbours of a transmuted site, impurities must be recognised and their class should not be updated. For mobile impurities, the N-Fold way algorithm should be adapted to account for IS and PU moves (see Appendix~\ref{app:NF} for details of the algorithm implementation), increasing the number of classes. In the impurity-free scenario, the computational effort introduced by additional code to handle variable updates in the N-Fold way algorithm is compensated by the time saved through the rejection-free transition to subsequent Markov states. This time efficiency improves as the temperature decreases, leading to the N-Fold way outperforming the Metropolis algorithm at low temperatures. For all the temperatures analysed in this study, the N-Fold way is the algorithm of choice in terms of CPU time efficiency.

To tune the dynamics for the N-Fold way algorithm (enhance or reduce the frequency of IS compared to PU), we modify Eq.~(\ref{eq:Glauber}) and introduce two different time constants $\tau_{IS}$ and $\tau_{PU}$ respectively for IS and PU moves that depend on $\alpha$. In this way, we modify the relative probability of selecting a PU or an IS move. The dynamics of the N-Fold way algorithm are constructed to match the Metropolis algorithm and, in particular, the acceptance ratio between IS and PU for the Metropolis algorithm needs to be equal to the selection ratio between the IS and PU for the N-Fold way algorithm. This condition is met if the time constants for IS and PU in Eq.~(\ref{eq:Glauber}) (respectively $\tau_{IS}$ and $\tau_{PU}$) satisfy the relation 
\begin{equation}
    \frac{\tau_{PU}}{\tau_{IS}}=\frac{\alpha}{(1-\alpha)}\frac{(1-f)}{4f}
    \label{eq:alpharatio}
\end{equation}
For a derivation, see Appendix \ref{app:ARE}. In our simulations, we consider without loss of generality $\tau_{PU}=1$ and derive $\tau_{IS}$ from Eq.~(\ref{eq:alpharatio}).

The time update $\Delta t$ between subsequent events of Eq.~(\ref{eq:dt_easy}) is modified to account for both PU and IS moves
\begin{equation}
    \Delta t = -\bigg(\frac{1-f}{1-\alpha}\bigg)\frac{\ln(r)}{\sum_i w_i^{PU}+\sum_j w_j^{IS}},
    \label{eq:dt}
\end{equation}
where
\begin{equation}
    w_i^{PU}(\Delta E_i^{PU}) = \frac{1}{2\tau_{PU}}\left(1-\tanh{\left(\beta \frac{\Delta E_i^{PU}}{2}\right)}\right),
    \label{eq:Glauber_PU}
\end{equation}
with $\Delta E_i^{PU}$ is the energy of the $i$-th class for the PU-group and 
\begin{equation}
    w_j^{IS}(\Delta E_j^{IS}) = \frac{1}{2\tau_{IS}}\left(1-\tanh{\left(\beta \frac{\Delta E_j^{IS}}{2}\right)}\right),
    \label{eq:Glauber_IS}
\end{equation}
with $\Delta E_j^{IS}$ is the energy of the $j$-th class for the IS-group.

To be noted, the only difference between the two algorithms is only related to time efficiency, as the average dynamical properties (e.g. time correlation functions, event rates) are the same independent of the algorithm considered as shown in section~\ref{sec:Results}.

\section{Classical Nucleation Theory} \label{sec:CNT}
In the context of solute precipitation, CNT describes the transition process from a metastable solution phase to a stable equilibrium phase through the birth and subsequent growth of nucleating droplets of solute particles. CNT is a well-accepted theory that calculates the nucleation rate based on a few assumptions. It assumes that droplets grow (shrink) through attachment (separation) of single solute particles and the nucleation is a one-step process with no intermediate phase formation. The key physical quantity that describes the process is the free energy function. The nucleation-free energy can be expressed as
\begin{equation}
    F(\lambda) = -B_1\lambda+A_1\sqrt{\lambda}+A_2\log(\lambda)+A_3
    \label{eq:F}
\end{equation}
where $\lambda$ is the cluster size, i.e., the number of solute particles in the droplet. We set the reference point of the free energy at the lowest cluster size  ($\lambda = 1$) to $F(1)=-k_BT\log[\rho_1(\Delta\mu, T)]$, where $\rho_1(\Delta\mu, T)$ is the monomer density, i.e. the density of isolated solutes at given chemical potential difference and temperature. As the metastable supersaturated state contains a negligible quantity of non-monomer solutes, this is an excellent approximation. The free energy expression is valid only for $\lambda \geq 1$ at any non-zero temperature. This form of the free energy was verified numerically by Ryu and Cai~\cite{Ryu2010}. The first term in the free energy expression accounts for the solute particles in the bulk of the droplet, and since it is an energy gain, we can approximately write $B_1=\Delta\mu$ at low temperatures. The second term is the energy required to form a border separating solute and solution, and the energy cost can be written as $A_1=2\sqrt{\pi}\sigma$, where $\sigma$ is the surface tension. The third term was introduced by Langer~\cite{Langer1968}, and it is a logarithmic correction due to fluctuations at the surface and deviation from the circular shape of the droplet. It can be derived theoretically for a homogeneous system where it takes the value $A_2 = \frac{5}{4}k_BT$. In the following, we assume $A_2$ to be invariant for the model with impurities. The last term $A_3$ adjusts the height of the free energy function by a constant shift. In particular, we take $A_3 = -k_BT\log[\rho_1(\Delta\mu, T)]+B_1-A_1$. 

In this study, we consider the geometrical definition of a cluster, consisting solely of particles (or spins with a value of $+1$, in magnetic terminology). Previous studies \cite{Ryu2010,Schmitz2013} have demonstrated that near the critical temperature $T_c$, significant discrepancies with the predictions of CNT arise when using this geometrical cluster definition. In the high-temperature regime, the physical definition of clusters should be considered. Under this framework, a cluster is defined as a collection of particles (or $+1$ spins) connected by an active bond \cite{Coniglio1980,Swendsen1987}. The probability of a bond being active is given by $p_B(T) = 1 - \exp(-2J/k_B T)$. Since our analysis is limited at $T<1.6$ [$p_B(T=1.6) \approx 0.71$] no large discrepancies are expected between the two definitions. We therefore proceed to consider the geometrical definition of clusters. In our simulations, we follow the cluster size variations after each IS and PU move. To perform this efficiently, we have designed a new cluster identification algorithm, which is presented in Appendix~\ref{app:CIA}. This is one of the key ingredients which has greatly reduced the computational cost related to nucleation rate calculations in this study.

The free energy maximum with respect to $\lambda$ is known as the barrier height, and the corresponding value of $\lambda$  is called critical cluster size $\lambda_c$. The crucial and lengthy step in the nucleation process is crossing this free energy barrier. Due to thermal fluctuations, small droplets appear and dissolve until one becomes bigger than the critical cluster size. The supercritical droplet shows steady growth with time. We can derive the analytical expression for the critical cluster size from Eq.~(\ref{eq:F}) as
\begin{equation}
    \lambda_c = \left[\frac{A_1 + \sqrt{A_1^2+16B_1A_2}}{4B_1}\right]^2.
    \label{eq:Lc}
\end{equation}

The key quantity under investigation in the present work is the nucleation rate $I$, defined as the number of post-critical nuclei formation events occurring per unit time and unit area in a system. Following BDZ theory, we can write a theoretical expression for the nucleation rate as
\begin{equation}
    I^{BD} = D_c\Gamma \exp{\left[-\frac{F(\lambda_c)}{k_BT}\right]},
\end{equation}
where $D_c$ is the diffusion coefficient of particles at the boundary and $\Gamma$ is the Zeldovich factor \cite{Zeldovich1943}.

The diffusion coefficient estimates the rate at which solute particles are attached at the boundary of a critical cluster. $D_c$ can be derived in simulations by preparing a critical cluster in equilibrium with the environment and following its dynamics. If $\lambda(t)$ is the size of the cluster at time $t$ that was prepared to have critical size $\lambda_c$ at time $t=0$, then
\begin{equation}
    D_c = \frac{\langle (\lambda(t)-\lambda_c)^2 \rangle }{2t},
    \label{eq:DC}
\end{equation}
where $\langle x\rangle$ represents the average of quantity $x$ over many realisations.

The Zeldovich factor captures the effect of free energy curvature near the critical size. It says that if two free-energy landscapes have equal barrier heights, the nucleation rate is slower for the one with flatter geometry near the maximum due to increased propensity for recrossing. The Zeldovich factor can be expressed as
\begin{equation}
    \Gamma = \frac{1}{\sqrt{2\pi k_BT}}\left[-\frac{\partial^2F(\lambda)}{\partial \lambda^2}\bigg{|}_{\lambda = \lambda_c}\right]^{1/2}.
\end{equation}

The nucleation rate can be estimated if we know the free energy landscape or the free energy as a function of cluster size. For this objective, we use umbrella sampling. An alternate and independent way to derive the nucleation rate is forward flux sampling. We briefly introduce both techniques in the next section.

\section{Rare-event sampling methods} \label{sec:RESM}
\subsection{Umbrella Sampling}\label{sec:US}
The free energy of a cluster of size $\lambda$ can be written as
\begin{equation}
    F^{US}(\lambda)=-k_BT\ln{(\rho_1)}-k_BT\ln\left[{\frac{P(\lambda)}{P(1)}}\right],
\end{equation}
where $\rho_1$ is the monomer density (density of isolated solute particles) in the parent phase and $P(\lambda)$ represents the probability that a given cluster has size $\lambda$. It can be written mathematically as
\begin{equation}
P(\lambda)=\frac{n(\lambda)}{\sum_{\lambda=1}^{\lambda_m}n(\lambda)},
\end{equation}
where $n(\lambda)$ is the average number of clusters of size $\lambda$ and $\lambda_m$ is the maximum cluster size that are sampled.
Note that the minimum size of a cluster $\lambda=1$ is our reference point. A reliable estimation of $P(\lambda)$ well beyond the critical cluster size is required to calculate the free energy barrier. This means a large range of $\lambda$ should be sampled (typically from $\lambda=1$ up to $\lambda\sim 10^3$). 

The US technique is employed to obtain the free energy as a function of the cluster size as a reaction coordinate. The total range of $\lambda$ is divided into windows with strict boundaries $[\lambda^n_i, \lambda^n_f)$, where $n$ represents the $n$-th window. The window size $\Delta \lambda$ and the overlap between two subsequent windows $\lambda_{over}^{n,n+1}$ can be changed for different temperatures and windows. The system is prepared with a cluster $\lambda \in [\lambda^n_i, \lambda^n_f)$ and simulated long enough subject to an infinite square well potential, which prevents the cluster size from leaving the window. In this work the simulation time was set to $5\times 10^3$ time units. Each move that overshoots $\lambda$ outside the window's boundaries is rejected, restoring the previous configuration. We derive the relative probability $p_n(\lambda)$ to sample a certain cluster size in the  $n$-th window as the ratio between the time spent by the system in such configuration and the total simulation time. Then
\begin{equation}
    P(\lambda) = \prod_{\lambda'=2}^{\lambda} \langle r_n(\lambda')\rangle_n = \prod_{\lambda'=2}^{\lambda}\left\langle \frac{p_n(\lambda')}{p_n(\lambda'-1)}\right\rangle_n,
\end{equation}
where $\langle r_n(\lambda') \rangle_n$ represents the mean over all the windows in which $\lambda' \in [\lambda'^n_i, \lambda'^n_f)$. 

When $F(\lambda)$ has a positive slope, the system spends more time near the left boundary of the window, which means that $p_n(\lambda)$ is more accurate for states with $\lambda$ near $\lambda^n_i$ and as $\lambda$ goes towards $\lambda^n_f$ the uncertainty increases. Setting $\Delta \lambda$ to a small value ensures that the US does not collect noisy data near the right boundary of the window $\lambda=\lambda^n_f$. A large value of $\lambda_{over}^{n,n+1}$ ensures that many independent measurements of the ratio $p_n(\lambda)/p_n(\lambda-1)$ are taken across multiple windows. Less noisy data and more measurements of the same ratio should provide better estimates of $F(\lambda)$, especially where the free energy $F(\lambda)$ is particularly steep. These tricks are more effective than simulating for a longer time with a lower number of windows and/or smaller overlap regions.

When sampling larger cluster sizes (where $F(\lambda)$ is less steep) $\lambda_{over}^{n,n+1}$ is reduced to decrease the number of windows and, therefore, reduce the computational cost. Table \ref{tb:lambda0} shows the minimum and maximum window length and the window's overlap, depending on the simulated temperature.

\subsection{Forward Flux Sampling}
The formation of a post-critical nucleus at low temperatures and chemical potential difference can be seen as a rare event. Its direct observation can easily exceed the available simulation time. For this reason, a different strategy must be adopted to derive the nucleation rate. One efficient technique is the FFS~\cite{2009_ffs_allen,2009_tps_escobedo,2005_ffs_allen,polymer_folding_allen_2012}, which decomposes the rare event into many smaller events with a higher probability of occurrence. The total probability is then the product of the probabilities of all connected single events. We divide the configuration space into a specific number of interfaces defined by the maximum cluster size in the simulation, up to the largest post-critical nucleus size we aim to simulate. The $i$-th interface contains all the configurations with clusters of size $\lambda_{i} = (i-1) d\lambda$ and $\lambda_{i+1} = i d\lambda$. We denote $P(\lambda_{i+1}|\lambda_{i})$ the probability of growing a cluster from $i$-th interface to $(i+1)$-th interface. The nucleation rate can be estimated by the expression
\begin{equation}
    I = \phi_0 \sum_{i=0}^NP(\lambda_{i+1}|\lambda_{i}),
    \label{eq:I}
\end{equation}
where $\phi_0$ is defined as the initial flux representing the number of clusters reaching the size $\lambda_0$ per unit time and area starting from the metastable phase. The final interface $N$ is chosen such that $\lambda_{N+1}$ is big enough from which the probability of forming a post-critical nucleus is $1$, i.e. $P(\lambda_{N+1}|\lambda_{N})=1$. This means the nucleus always grows from cluster size $\lambda_{N+1}$. The transition probability $P(\lambda_{i+1}|\lambda_{i})$ is derived by sampling (from simulations at the previous interface) 960 cluster configurations with size $\lambda_{i}$ and counting the number of clusters that reach size $\lambda_{i+1}$ before shrinking to size $\bar{\lambda}$, which is set to be below $\lambda_0$ (see Table \ref{tb:lambda0} for values).

\subsection{Averaging random configurations of impurities}
Since different random configurations of impurities affect the nucleation rate calculation for static impurities, it is important to average over the impurity realisations. Let's consider $n^2$ realisations, each with nucleation time $\tau_i$ and nucleation rate $I_i=\tau_i^{-1}L^{-2}$. Every realisation can be seen as a tile of a larger system with size $L_{tot} = nL$. We aim at reconstructing the nucleation rate $I_{tot}$ of the whole system from its constituents $I_i$. We will neglect the possibility of nucleation across more than one tile, which is reasonable for large enough sub-systems. Considering the nucleation rate as the number of events per unit of time and area, it is easy to show that
\begin{equation}
I_{tot} =\frac{1}{ \tau_{tot} L_{tot}^2} = \sum_i\frac{1}{\tau_i} \frac{1}{n^2L^2} = [I_i]
\end{equation}
where $[ \dots ]$ indicates the average over randomness realisations and $\tau_{tot}$ the total nucleation time event for the whole system. This easily follows once we consider that the nucleation of a droplet is a Poissonian process. Indeed, the expectation time for a nucleation event in the whole system $\tau_{tot}$ can be derived as the harmonic mean of the expectation time $\tau_i$ of $n^2$ subsystems.

For the BDZ theory, the average over randomness should be done after deriving the nucleation rate $I_i$ for each impurity realisation. Therefore, the estimation of the diffusivity coefficient, Zeldovich factor and free energy barrier are necessary for each impurity realisation.

\section{Results} \label{sec:Results}
\begin{figure}
    \centering
    \includegraphics[width=0.8\columnwidth]{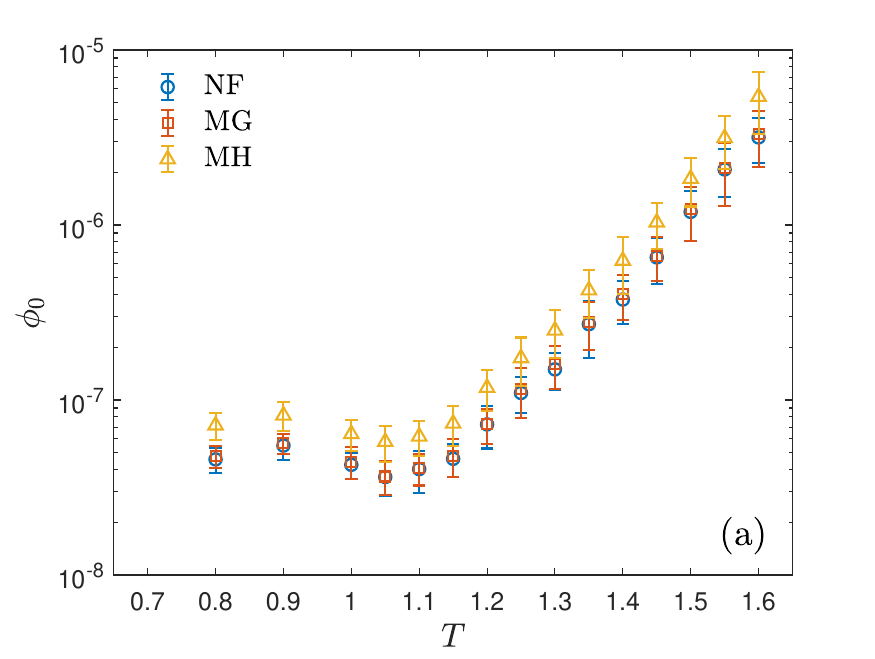} 
    \includegraphics[width=0.8\columnwidth]{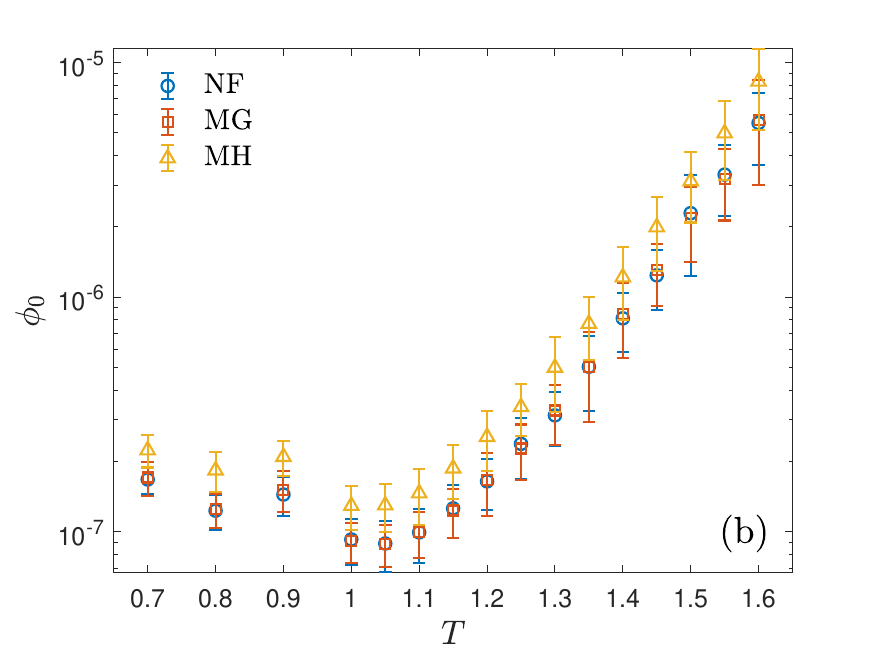} 
    \caption{Initial flux as a function of temperature, calculated with different algorithms for static and dynamic impurities, panel (a): static impurities; panel (b): dynamic impurities at fixed $f=2\%$ and $\Delta \mu=0.1$. Here, NF stands for N-Fold way algorithm with Glauber dynamics, MG for Metropolis with Glauber dynamics, MH for Metropolis-Hastings. The threshold values of cluster size $\lambda_0$ used in initial flux calculations are reported in Table~\ref{tb:lambda0}. $\lambda_0$ is chosen to have sampling probability lower than $0.99$ in unbiased simulations.}
    \label{fig:IF}
\end{figure}

We start showing the equivalence of the dynamics between the N-Fold way algorithm and the Metropolis algorithm. Figure~\ref{fig:IF}(a) depicts the initial flux $\phi_0$ as a function of temperature $T$ for a system with $f=2\%$ of static impurities at $\Delta\mu=0.1$. The initial flux is calculated as the number of crossings of the cluster size threshold $\lambda_0$ per unit time and unit area, starting from the metastable phase. We change the value of $\lambda_0$ with temperature to gain time efficiency. The values of $\lambda_0$, used in the simulations are reported in table~\ref{tb:lambda0}. The initial flux $\phi_0$ is obtained over $6400$ crossings for each data point and the error bars are computed from the standard errors in the distributions of times between crossing events. The results obtained from the N-Fold way (NF) and the Metropolis-Glauber (MG) algorithms show nice agreement throughout the temperature range plotted in Fig.~\ref{fig:IF}(a). However, a consistent difference can be seen while comparing the results obtained by the Metropolis-Hastings (MH) algorithm. The difference could be attributed to the dynamic dissimilarity (Glauber in the N-Fold way case versus Metropolis). The estimated values of initial flux are almost $1.5$ times higher using MH throughout the temperature range studied. Here we see non-monotonic behaviour in the $\phi_0$ vs. $T$ plot, which arises from using different $\lambda_0$ values at different temperatures. This choice is necessary to accommodate varying simulation conditions: $\lambda_0$ must be large enough to ensure that initial crossings of the threshold value to determine $\phi_0$ are rare, but not so large that it increases waiting times excessively. For example, $\lambda_0 = 3$ is suitable for the lowest temperature $T = 0.7$ but too small for the highest temperature $T = 1.6$, while $\lambda_0 = 19$ is appropriate for the highest temperature but impractical for the lowest temperature. Fixing $\lambda_0$ across all temperatures would eliminate the non-monotonic behaviour. Dynamic impurities also give similar results as shown in Fig.~\ref{fig:IF}(b). We also notice that the variation in the impurity mobility doesn't change the qualitative results, although increasing the fraction of impurities enhances the initial flux.

\begin{figure}
    \centering
    \includegraphics[width=0.8\columnwidth]{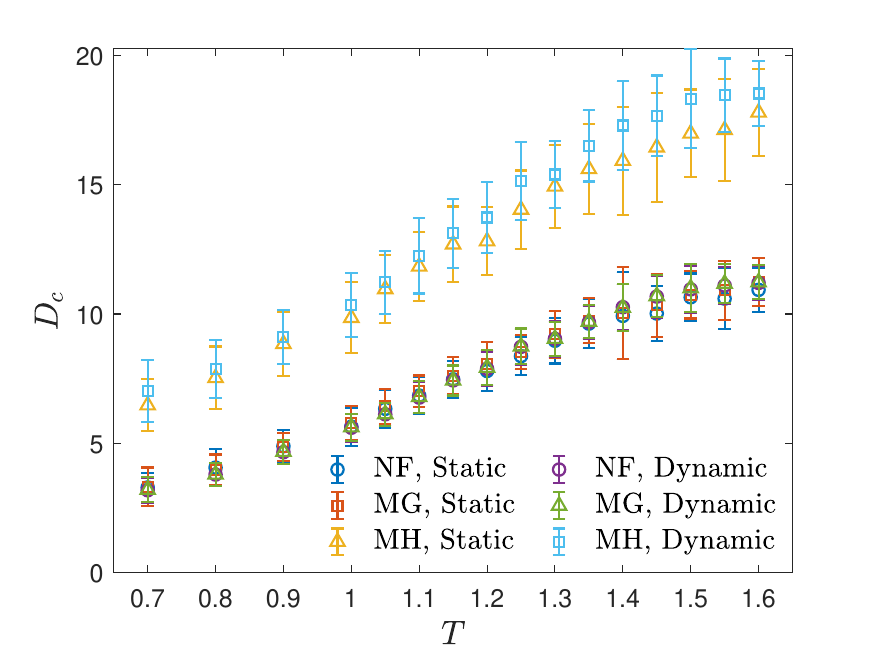} 
    \caption{Impurity diffusion coefficient as a function of temperature, computed considering different algorithms for static and dynamic impurities, for $f=2\%$ and $\Delta\mu=0.1$. NF stands for N-Fold way algorithm with Glauber dynamics, MG for Metropolis with Glauber dynamics, MH for Metropolis-Hastings. Error bars are obtained considering $10$ independent measurements of the coefficient, averaging over $10^4$ trajectories.}
    \label{fig:DC}
\end{figure}

Another dynamic-dependent parameter is the diffusivity coefficient $D_c$. In Fig.~\ref{fig:DC}, $D_c$ [see Eq.~(\ref{eq:DC}) for definition] as a function of $T$ is plotted for both static and dynamic impurities respectively. For each data point, we record $10^5$ trajectories of the time-evolution up to 1 physical time unit given by Eq.~(\ref{eq:dt}) (equivalent to 1 MCSS for the Metropolis algorithm) for a system prepared with the largest cluster with critical size. Then we obtain independent measurements of $D_c$ considering $10$ batches of $10^4$ trajectories. The data point in Fig.~\ref{fig:DC} is the average of these values and the error bars are their standard deviation. For the diffusivity coefficient, both NF and MG give approximately similar estimates. However, a difference with MH is observed, with MH values being $1.8$ times higher than MG values. We also observe that $D_c$ decreases monotonically with decreasing temperature, in accordance with previous studies~\cite{Mandal2021}.

\begin{table}
\begin{center}
\begin{tabular}{|c |c c c c c|}
 \hline
 $T$ & $\lambda_0$ & $\bar{\lambda}$ & $\lambda_{over}^{min}$ & $\lambda_{over}^{max}$ & $\Delta \lambda$ \\ 
 \hline
 1.6 & 19 & 7 & 10 & 18 & 20 \\
 1.55 & 18 & 6 & 10 & 18 & 20 \\
 1.5 & 17 & 6 & 10 & 18 & 20 \\
 1.45 & 16 & 6 & 10 & 18 & 20 \\
 1.4 & 15 & 6 & 10 & 18 & 20 \\
 1.35 & 14 & 6 & 10 & 18 & 20 \\
 1.3 & 13 & 5 & 10 & 18 & 20 \\
 1.25 & 12 & 5 & 10 & 18 & 20  \\
 1.2 & 11 & 5 & 10 & 18 & 20 \\
 1.15 & 10 & 5 & 10 & 18 & 20 \\
 1.1 & 9 & 4 & 5 & 8 & 10 \\
 1.05 & 8 & 4 & 5 & 8 & 10 \\
 1 & 7 & 4 & 5 & 9 & 10 \\
 0.9 & 5 & 3 & 5 & 9 & 10 \\
 0.8 & 4 & 3 & 5 & 9 & 10 \\
 0.7 & 3 & 2 & 5 & 9 & 10 \\
 \hline
\end{tabular}
\end{center}
\caption{Simulation parameters, employed at different temperatures for the FFS and US calculations. $\lambda_0$ is the threshold cluster size for the initial flux calculation; $\bar{\lambda}$ corresponds to the minimum cluster size below which a trajectory is considered as a failed attempt of reaching to the next interface in the FFS simulation; $\lambda_{over}^{min}$ ($\lambda_{over}^{max}$) is the minimum (maximum) overlap between subsequent windows in the US calculations; $\Delta \lambda$ is the window length used for US calculations.}
\label{tb:lambda0}
\end{table}

\subsection{Efficiency in Computational Time} \label{sec:Results_CPU}
\begin{figure}
    \centering
    \includegraphics[width=0.8\columnwidth]{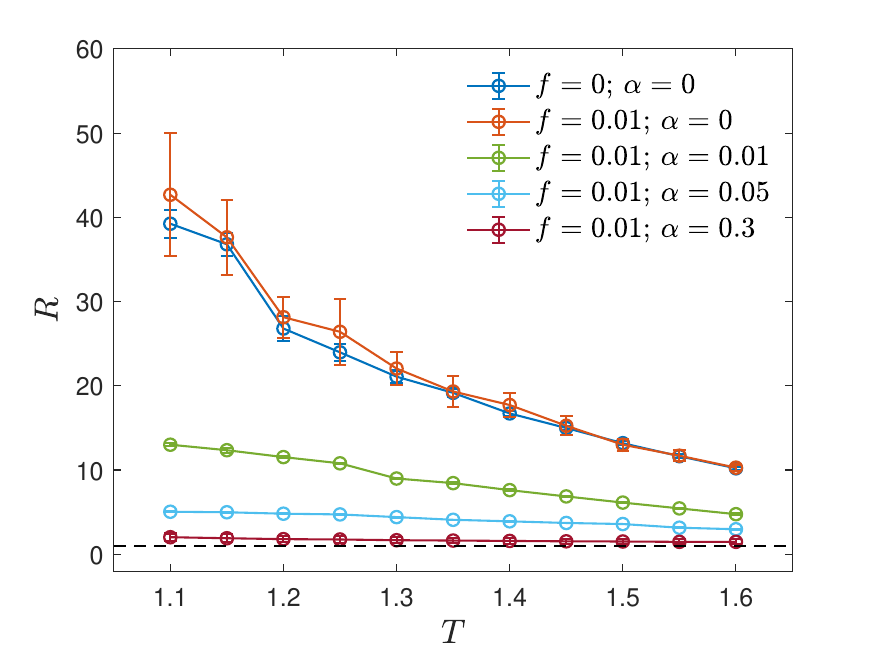}  
    \caption{CPU time ratio between the Monte Carlo algorithm and the N-Fold way algorithm, as a function of temperature, for different chemical potential differences, fraction of impurities and mobility. The two algorithms have the same cluster recognition and counting algorithm, presented in Appendix \ref{app:CIA}. Chemical potential difference $\Delta\mu$ is fixed to $0.1$ for all the cases. The dashed line represents $R=1$, where the Metropolis and N-Fold way algorithms perform equally.}
    \label{fig:NFvsM_Comparison}
\end{figure}

We now compare the efficiency between the Metropolis algorithm with Glauber dynamics and the N-Fold way algorithms for different temperatures. As expected, the nucleation rates obtained using the two algorithms are equal within statistical confidence. The ratio $R = t^{MG}/t^{NF}$ of the CPU execution time for the computation of the nucleation rate of the two algorithms is plotted in Fig.~\ref{fig:NFvsM_Comparison} for three different cases: no impurities, static impurities and dynamic impurities. For the first two cases, the time gain reached a factor of 40 at temperature $T=1.1$. Even though the execution time of a simulation strongly depends on its optimisation, this dramatic improvement in the simulation time is related to the choice of the algorithm used. The temperature dependence shows that, at low temperatures, the preferred choice of algorithm is the N-Fold way. However, the enhancement in $R$ for mobile impurities is less dramatic, especially when the mobility parameter of impurities $\alpha$ increases. This effect is due to the increased computational effort per move in the N-Fold way algorithm when considering mobile impurities, particularly while an IS move is selected.

The \emph{growth} in $R$ with decreasing temperature is also much less pronounced in the case of dynamic impurities. To understand this, let us consider the ratio between accepted PU and IS moves. Since the system dynamics are the same, we can switch to the Metropolis framework instead of working with the N-Fold way one for the following derivation. Most IS moves have null energy costs at both high and low temperatures and will be accepted with probability given by Eq.~\ref{eq:Glauber} $w_G(\Delta E=0) = 0.5$. Indeed, our simulations confirm that most impurities navigate into solvent or solution, and IS moves have a lower impact on PU moves in changing the system's free energy. Conversely, PU moves are strongly temperature-related. At low temperatures, in both parent and child phases, PU moves require a high energy cost $\Delta E \approx 8\pm \Delta\mu$ and $w_G(\Delta E)$ sharply decreases with decreasing temperature. Hence, even though $\alpha<0.5$, most of the accepted moves will be IS rather than PU at low temperatures. Our simulations shows that $A_{IS}/(A_{IS}+A_{PU})> 0.60$ at $T=1.5$ and $> 0.95$ at $T=1$, for $\alpha>0.05$. The ratio $A_{IS}/(A_{IS}+A_{PU})$ quickly saturates as temperature is lowered. Since IS moves do not contribute substantially to completing the nucleation process, the CPU time ratio between the two algorithms at low temperatures is simply equal to the ratio between the time taken to accept an IS move (for the Metropolis) and the time to update an IS move (for the N-Fold way).

\subsection{Nucleation free energy}
\begin{figure}
    \centering
    \includegraphics[width=0.8\columnwidth]{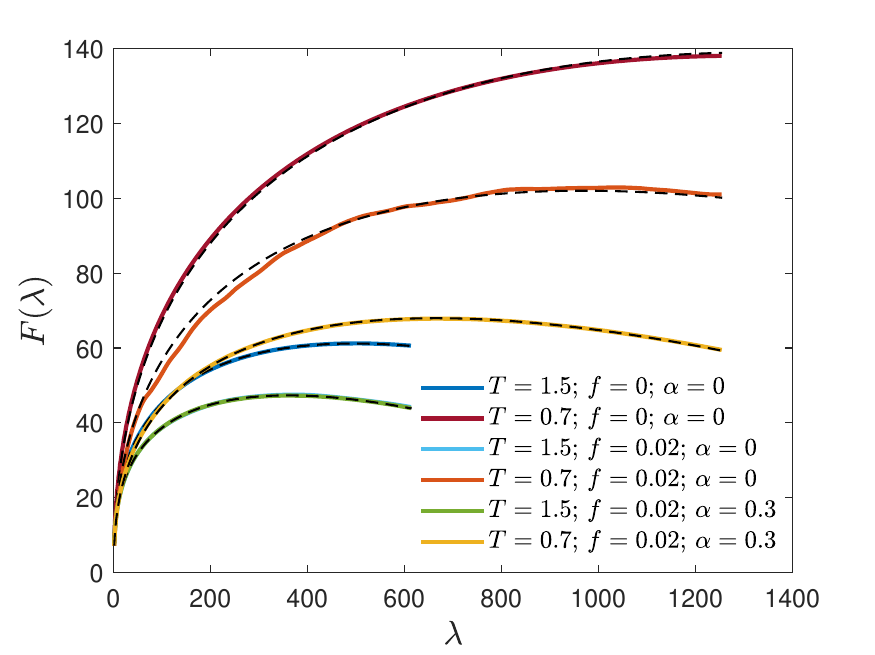}
    \caption{Free energy as a function of cluster size for different temperatures and impurity fractions. Results from the US techniques are shown in solid lines and the fit with Eq.~(\ref{eq:F}) is shown in dotted lines. For $T=1.5$, the two curves for static and dynamic impurities are superimposed as no major difference can be noticed in such conditions.}
    \label{fig:US_FreeEnergy}
\end{figure}

In the rest of this paper, we adopt the N-Fold way algorithm for all Monte Carlo simulations. The free energy of the nucleation process is derived using the US technique, as described in section~\ref{sec:US}. We fit the free energy to the function shown in Eq.~(\ref{eq:F}), considering the interfacial tension ($A_1$) as a fitting parameter in the case of static impurities. However, in the case of dynamic impurities, both interfacial tension ($A_1$) and bulk free energy coefficient ($B_1$) are considered fitting parameters. These are the quantities that can be potentially influenced by the presence of mobile impurities. Figure~\ref{fig:US_FreeEnergy} shows the US results along with the fit for different temperature values, impurity fractions and mobility coefficients. Due to limitations in simulation time, we could not derive the free energy curves up to the critical cluster size for the no impurity case at low temperatures. Using the fitting parameters, we can derive the critical cluster size $\lambda_c$ and the free energy barrier $F(\lambda_c)$ via Eq.~(\ref{eq:Lc}) and Eq.~(\ref{eq:F}), respectively.

It must be noted that, in the case of heterogeneous nucleation, the cluster size obtained through the nearest-neighbour connectivity may not be the proper reaction coordinate to describe the nucleation process. As found by Yao et al.~\cite{Yao2023}, for the three-dimensional random field Ising model, the location of the first nucleus should also be considered, especially when the randomness increases in the system. We find similar effects for the static impurity case, for which the trajectories in the US calculations are impurity realisation-dependent. However, due to the low impurity fractions, we use the cluster size as a reaction coordinate for the systems studied in the paper. We do an average over many impurity realisations to minimise statistical fluctuations. In the case of mobile impurities, the space symmetry is restored, and the problem of reaction coordinate does not arise.

\begin{figure}
    \centering
    \includegraphics[width=0.8\columnwidth]{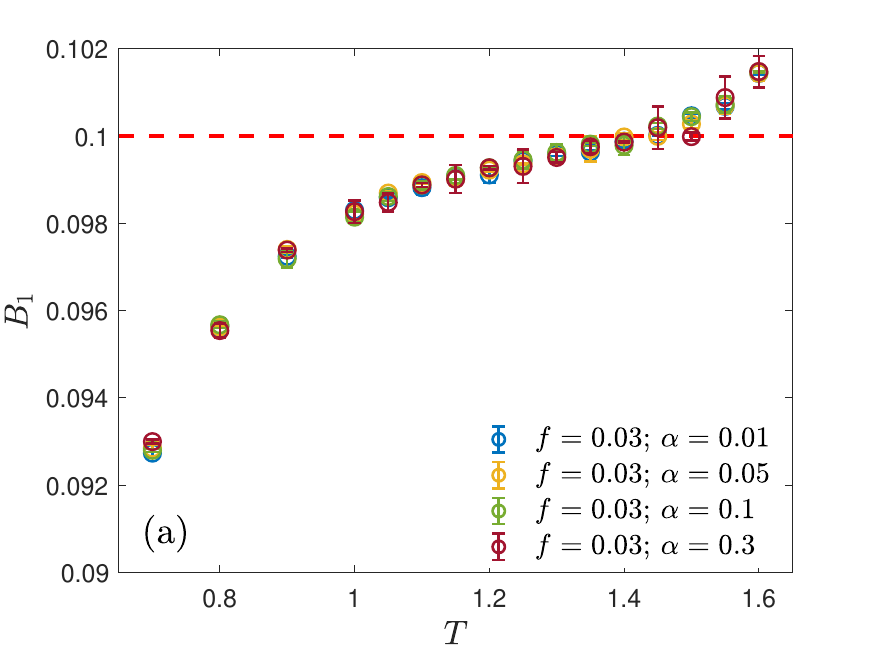}
    \includegraphics[width=0.8\columnwidth]{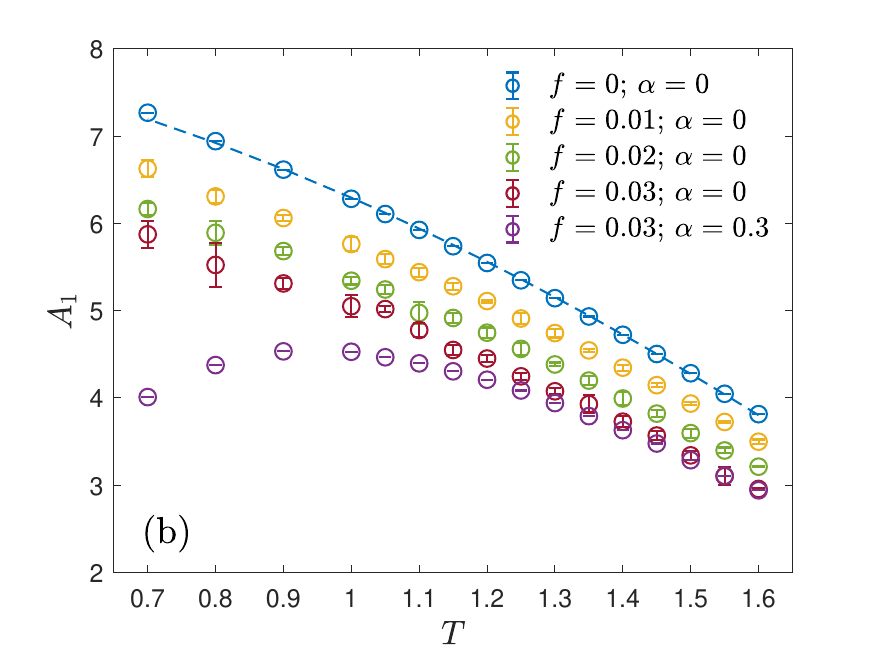}
    \caption{Fitting parameters for the US simulations for a system with no impurities, static impurities and mobile impurities. Panel (a) shows the estimation of $B_1$ written in Eq.~(\ref{eq:F}) for the mobile impurity case, along with the reference value $\Delta\mu=0.1$ shown by a dashed line. Panel (b) shows the estimation of $A_1$, which is proportional to the surface tension. The dashed line has been theoretically derived for the case of no impurities~\cite{Shneidman1999}.}
    \label{fig:US02}
\end{figure}

Figure~\ref{fig:US02} summarises our findings regarding the fitting parameters in Eq.~(\ref{eq:F}). Panel (a) shows the bulk coefficient $B_1$ as a function of temperature for the case of mobile impurities with impurity fraction $f=0.03$. As specified earlier, for the homogeneous and static impurity cases, the bulk term is fixed to $\Delta\mu$, here shown by the dashed line. For high temperatures, small (but statistically relevant) deviations from the theoretical expression can be observed. This effect could be related to the geometrical cluster assumption, which does not hold for temperatures near the critical temperature $T_c$~\cite{Binder2016}. The physical cluster definition should be considered in this regime, which gives smaller cluster sizes. If we instead consider the geometrical clusters, the fitted effective chemical potential difference from Eq.~(\ref{eq:F}) produces a lower value with respect to the reference to compensate for larger cluster areas. At low temperatures, the deviation of $B_1$ from the reference $\Delta\mu$ is more substantial. This could be because of preferential bonding between impurity and solute at low temperatures. Similar behaviour is observed in the case of non-zero, anti-symmetric impurity-solute and impurity-solvent interactions when $B_1$ is calculated at different values of interaction strength~\cite{Mandal2024}. The dotted line in panel (b) represents the analytical expression of the interfacial tension~\cite{Shneidman1999}, which agrees well with our analysis. The presence of impurities lowers the interfacial tension, especially for mobile impurities, which confirms the findings by Mandal and Quigley~\cite{Mandal2021}. In the case of mobile impurities, the interface free energy lowers further due to a gathering of impurities at the boundary of clusters (enhancing the screening effect between the stable and parent phases), especially at low temperatures. 

\begin{figure}
    \centering
    \includegraphics[width=0.8\linewidth]{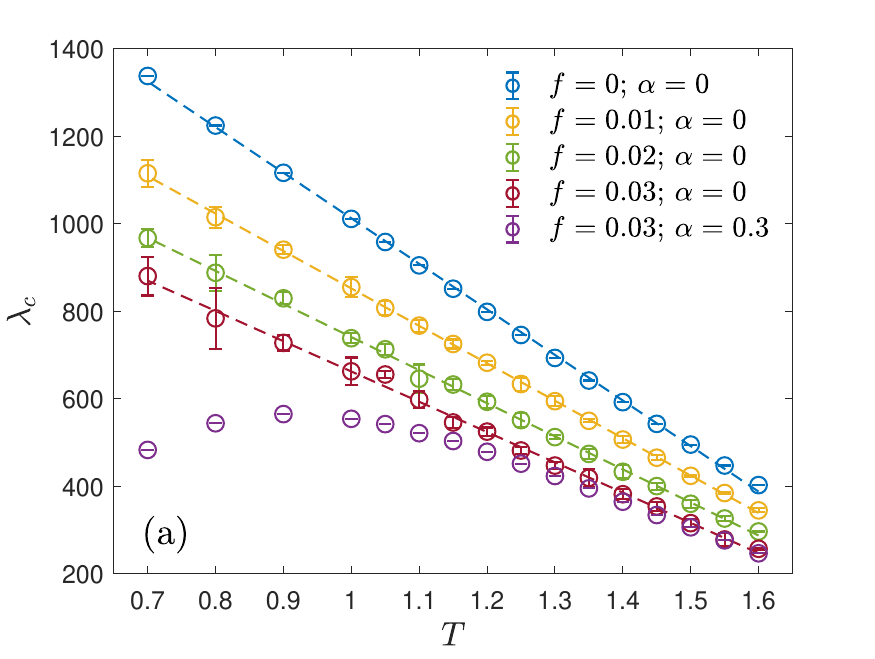}
    \includegraphics[width=0.8\linewidth]{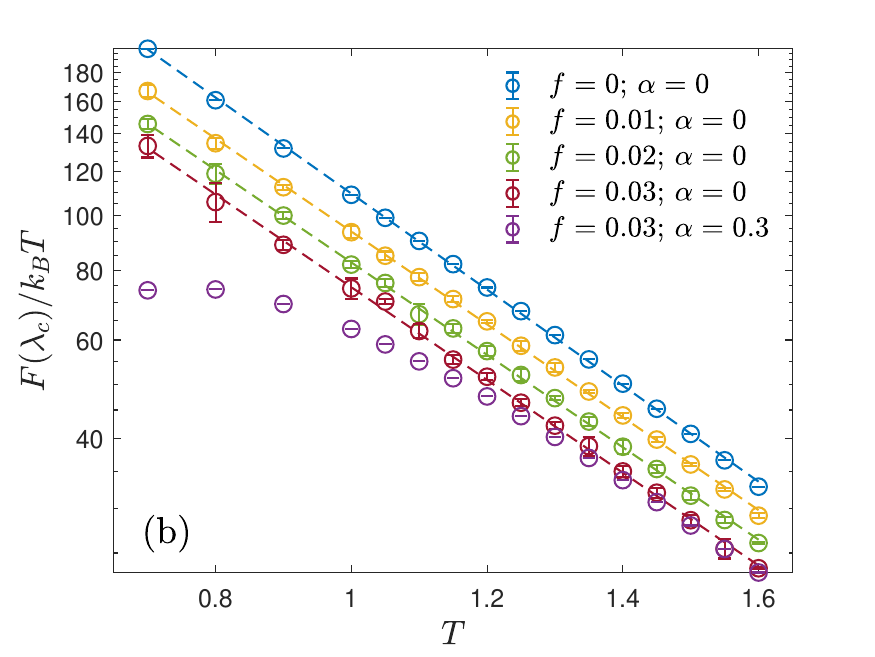}
    \caption{US estimation for a system with no impurities, static and mobile impurities. Panel (a) shows the estimation of the critical cluster. A linear fit can be superimposed for no and static impurities (dashed lines). Panel (b) shows the free energy barrier in $k_B T$ units in a semi-logarithmic plot. A linear fit can be superimposed for the case of no and static impurities (dashed lines).}
    \label{fig:US03}
\end{figure}

Panel (a) of Fig.~\ref{fig:US03}  shows the critical cluster size as a function of temperature. Interestingly, for static and impurity-free cases, linear behaviour can be observed. Panel (b) shows the free energy barrier and temperature ratio in a semi-logarithmic plot. A linear interpolation can be fit but ceases to exist (as also for the critical cluster size case) for mobile impurities in the low-temperature regime where the gathering of impurities at the cluster's boundary becomes substantial. For $T<0.9$, the scaled free energy barrier height $F(\lambda_c)/k_BT$ seems to saturate with $T$.

\subsection{Nucleation rate}
\begin{figure}
    \centering
    \includegraphics[width=0.49\columnwidth]{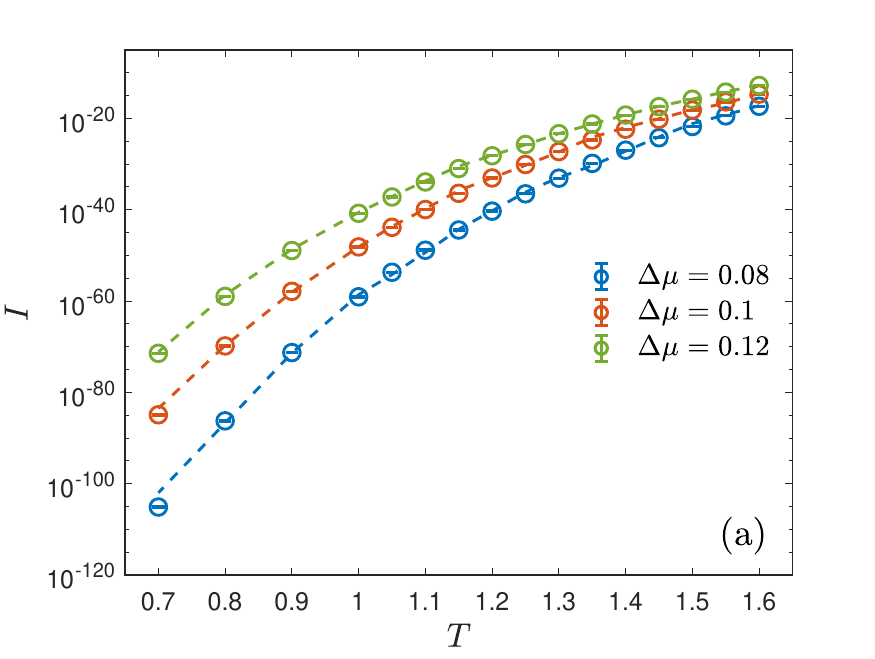}
    \includegraphics[width=0.49\columnwidth]{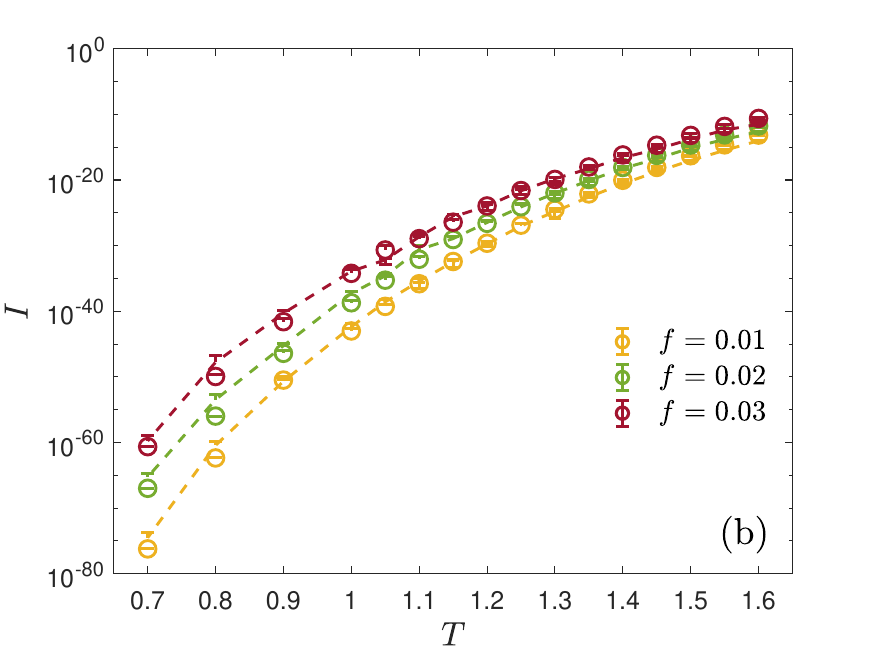}
    \includegraphics[width=0.8\columnwidth]{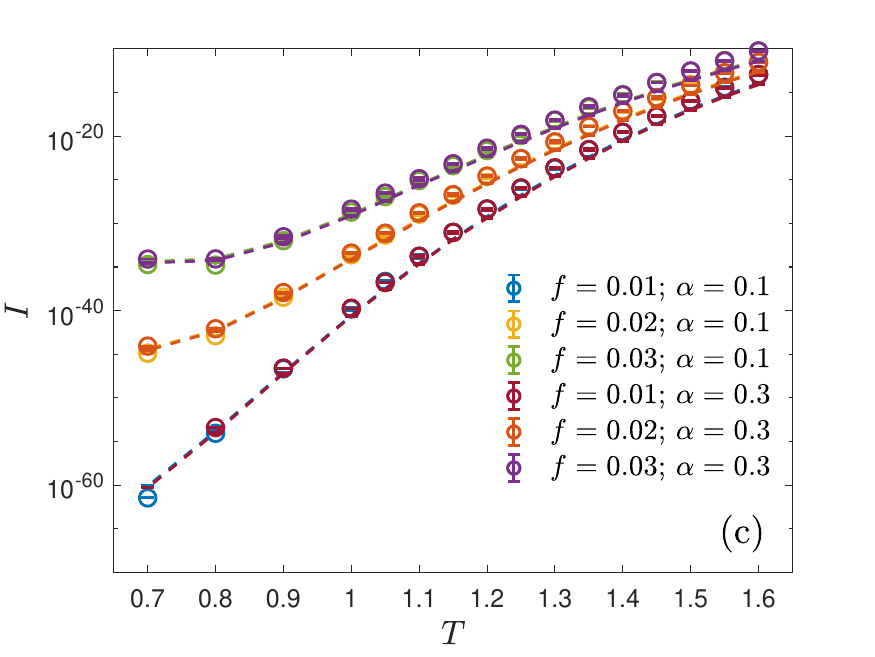}
    \caption{Nucleation rate as a function of temperature, for different chemical potential differences, impurity fraction and mobility (for the model with moving impurities) derived from FFS procedure (points) and the BDZ theory (dotted lines). Panel (a) shows the nucleation rate for different chemical potential differences in the absence of impurities. To be noted, $I \sim 10^{-100}$ was found for the lowest $\Delta\mu$ and temperature, which is $50$ orders of magnitude smaller than other low-temperature estimates found in the literature~\cite{Ryu2010}. Panel (b) shows the nucleation rate estimation for $\Delta\mu=0.1$ and different impurity fractions. Panel (c) shows the nucleation rate estimation for $\Delta\mu=0.1$ and $f=0.01$ for different impurity mobility values. FFS and BDZ theory show good agreement in all the cases. The estimations for $\alpha=0.3$ and $0.1$ are practically superimposed.}
    \label{fig:NF_NucleationRate}
\end{figure}

We analyse the nucleation rate for a wide range of temperatures: $0.7 \leq T \leq 1.6$. This is a substantial improvement for the FFS technique, for which previous studies were limited only for $T \geq 0.44T_c$~\cite{Ryu2010,Mandal2021}, or equivalently for $T \geq 1$. Here, $T_c$ is the critical temperature for the two-dimensional Ising model. As shown in panel (a) of Fig.~\ref{fig:NF_NucleationRate}, the lowest estimated nucleation rate for the impurity-free case reaches $\approx 10^{-100}$, which is $50$ orders of magnitude smaller than other estimations found in the scientific literature \cite{Ryu2010,Mandal2021}. To our knowledge, this represents the lowest estimated nucleation rate value derived using Monte Carlo simulations for the lattice-gas model. The nucleation rates obtained using BDZ theory [see Eq.~(\ref{eq:I})] and US barrier heights agree with the direct estimations via FFS simulations for all impurity fractions, high mobility coefficient and temperature ranges studied in the paper. Interestingly, $\alpha=0.3$ and $\alpha=0.1$ cases give identical nucleation rates. This can be justified considering that, as already said in sec.~(\ref{sec:Results_CPU}), at low temperatures and for $\alpha > 0.1$ the system dynamics is predominantly composed of IS moves. Therefore, within this regime, impurities are always in local equilibrium with the evolving clusters, and no dependency of the nucleation rate on $\alpha$ is observed.

\begin{figure}
    \centering
    \includegraphics[width=0.49\linewidth]{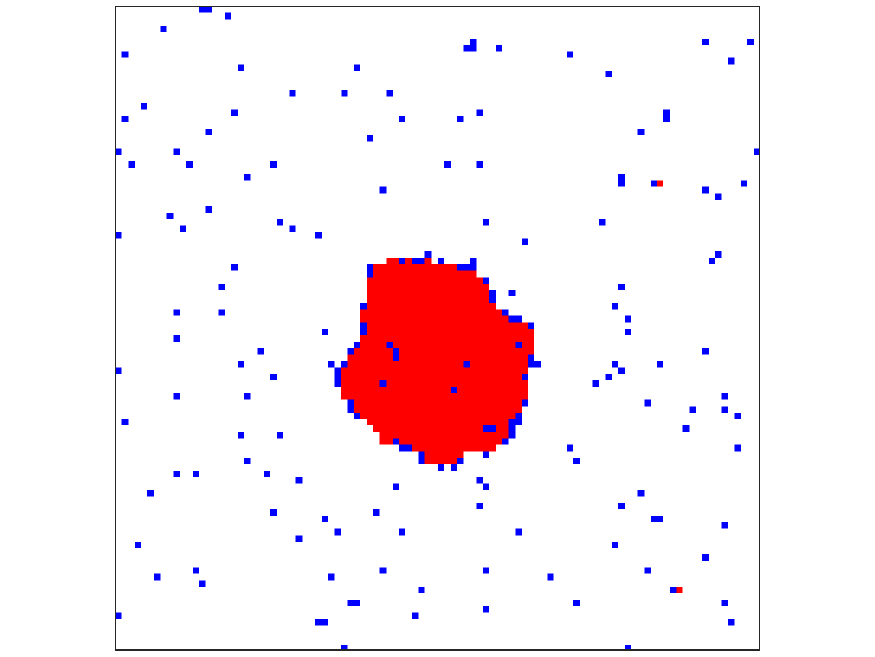}
    \includegraphics[width=0.49\linewidth]{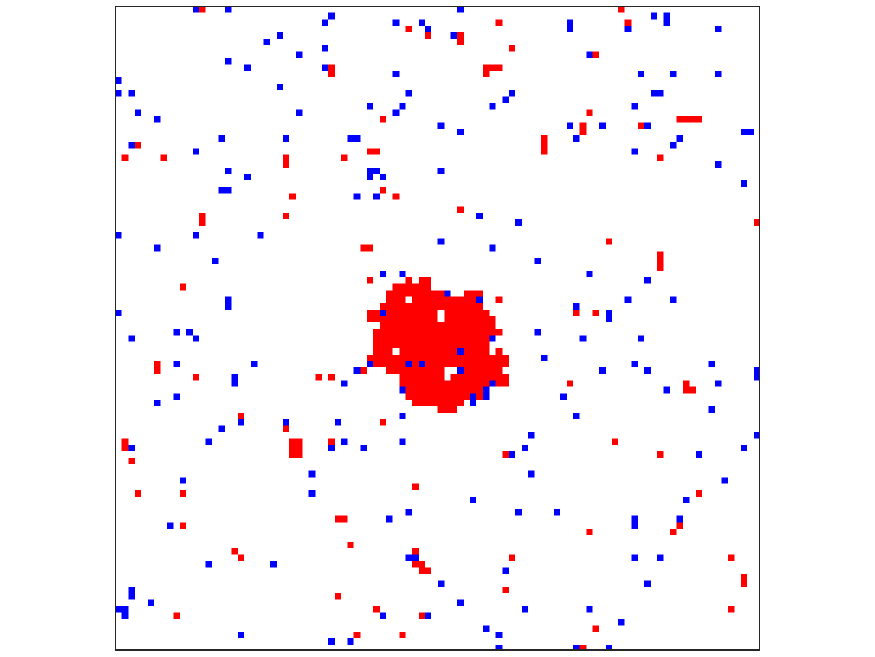}
    \caption{Snapshots of clusters near critical size in the US simulation for low impurity mobility ($\alpha=0.01$) at two different temperatures: $T=0.8$ (left) and $T=1.6$ (right). The cluster size and the fraction of defects at the boundary are respectively $\lambda_c=678$ and $\psi=0.52$ for the low-temperature case and $\lambda_c=299$ and $\psi=0.13$ for the high-temperature case. For these simulations, we have adopted $L=100$, $f=0.02$ and $\Delta \mu=0.1$. Impurities and particles are respectively denoted by blue and red colours, and white represents an empty site. At low temperatures, impurities accumulate at the cluster boundary, while at high temperatures, they are more dispersed.}
    \label{fig:HighLowPsi}
\end{figure}

By comparing panels (b) and (c) of Fig.~\ref{fig:NF_NucleationRate}, we can observe the effect of impurity mobility, which becomes more pronounced at lower temperatures when impurities have a stronger influence on the interfacial tension of a growing cluster. This effect is illustrated in Fig.~\ref{fig:HighLowPsi}, which shows snapshots of the cluster in critical conditions from US simulations for the low-mobility case ($\alpha=0.01$) at two different temperatures: $T=0.8$ (left) and $T=1.6$ (right). At lower temperatures, impurities accumulate more effectively at the cluster interface, leading to a reduction of the surface tension. This decrease lowers the free energy barrier for nucleation, thereby enhancing the nucleation rate, as observed in our results. Conversely, at higher temperatures, impurities become more dispersed, weakening their influence on nucleation dynamics. These findings are consistent with the observations reported by Mandal and Quigley~\cite{Mandal2021}.

\begin{figure}
\centering
\includegraphics[width=0.8\columnwidth]{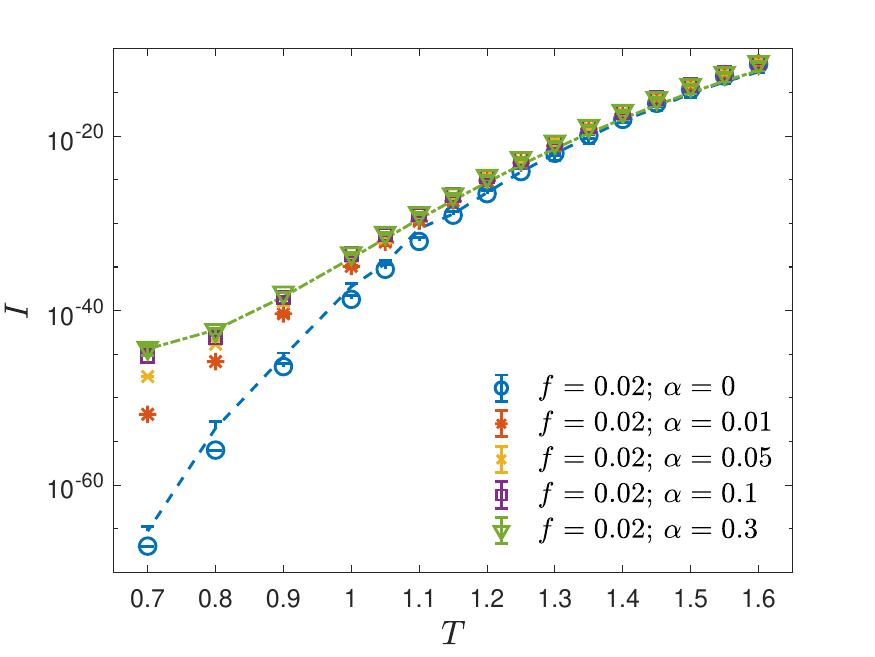}
\caption{Nucleation rate obtained from both BDZ theory using US (dotted lines) and FFS (markers) for $f=0.02$ and $\Delta\mu=0.1$. Different mobilities are represented, along with the static case. BDZ using US results follow FFS calculations for static and high-mobility cases. For the intermediate values of $\alpha$, instead, a discrepancy is observed between them. In particular, all BDZ calculations for mobile impurities are overlapped and in agreement with the high mobility FFS case.}
    \label{fig:NF_NRDynamic}
\end{figure}

An interesting scenario arises when we calculate the nucleation rate at low values of impurity mobility coefficient $\alpha$ as shown in Fig.~\ref{fig:NF_NRDynamic} for $f=0.02$.
Starting from $\alpha=0$ (static impurities) we increase $\alpha$ up to $0.3$. Symbols represent the nucleation rates obtained from FFS and lines correspond to the estimates from BDZ theory using US barriers. In the low-temperature regime, we observe a steady increase in nucleation rates, measured from the static ($\alpha=0$) to the mobile impurities with intermediate value ($\alpha=0.1$) mobility coefficient, using FFS. Beyond $\alpha=0.1$, we notice a saturation in the measured nucleation rates. However, this gradual shift in nucleation rate is not observed when we estimate rates using the US technique via BDZ theory. In this case, we always get saturated rates even for minimal non-zero values of $\alpha$. This mismatch in nucleation rates between FFS and BDZ theory at the intermediate values of $\alpha$ could be because of the timescale difference between cluster growth and impurity movement. For US simulations, as the cluster size is bound within a window, the nucleus quickly reaches a quasi-equilibrium state with mobile impurities for any non-zero $\alpha$, which is not possible in FFS simulations. In FFS there is no bound on cluster size and rapid transformation of clusters is observed while trajectories transition between interfaces. 

\begin{figure}
    \centering
    \includegraphics[width=0.8\linewidth]{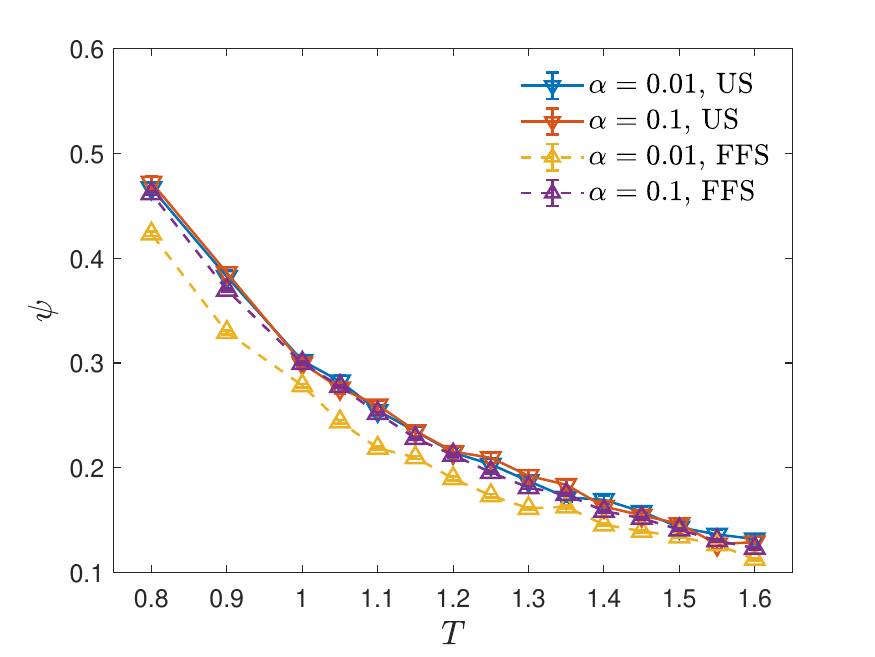}
    \caption{Fraction $\psi$ of boundary sites occupied by impurities as a function of temperature for US and FFS simulations, comparing low ($\alpha=0.01$) and high ($\alpha=0.1$) impurity mobilities. For all the analysed conditions, $\psi$ was determined for near-critical clusters. For these simulations $L=100$, $f=0.02$ and $\Delta \mu=0.1$. While US results do not depend on $\alpha$, FFS simulations show important differences between the two mobility cases.}
    \label{fig:Psi}
\end{figure}

To characterise this effect, we track the fraction $\psi$ of sites occupied by impurities at the boundary of a critical cluster. The fraction $\psi$ is mathematically defined as the number of impurity sites in direct contact with the critical cluster, divided by the total number of sites in contact with the cluster boundary.  Fig.~\ref{fig:Psi} presents $\psi$ as a function of temperature for both US and FFS simulations, considering two impurity mobilities: low ($\alpha=0.01$) and high ($\alpha=0.1$). As expected, in all temperature regimes, US simulations remain largely insensitive to impurity mobility, showing values of $\psi$ that coincide with the high-mobility case of FFS simulations. This agreement supports the idea that, in US simulations, impurities equilibrate rapidly with the cluster while it is restrained by the umbrella potential, independent of their intrinsic mobility. In contrast, FFS simulations show a mobility-dependent $\psi$, capturing the interplay between nucleation kinetics and impurity dynamics.

In conclusion, we report that CNT breaks down at very low temperatures and in the presence of mobile impurities. Since the impurity equilibrium condition during each step of the droplet growth process influences the nucleation process, the cluster size cannot be considered the only reaction coordinate. Another physical quantity expressing the relative condition between the droplet growth and the configuration of the impurities should be taken into account. A two-dimensional free energy function $F(\lambda,\phi)$ should then be analysed and the US technique accordingly adapted to produce a reliable estimation. We defer this to future work.

\section{Conclusion} \label{sec:Conclusion}
We have studied the nucleation properties of the two-dimensional Ising lattice-gas model in the presence of both static and dynamic impurities, specifically at low temperatures using the N-Fold way algorithm. The low-temperature regime has been of special interest as the surfactant properties of impurities can only be quantitatively studied at low temperatures, which were previously inaccessible.  Estimation of free energy barrier and nucleation rates is difficult at low temperatures ($T<1$) using the conventional Metropolis-Hastings or Metropolis-Glauber algorithm as most of the attempted update moves are rejected due to miniscule acceptance probability $e^{-\Delta E/k_BT}$. We need the N-Fold way algorithm to overcome this slowdown. We have shown that the gain in computational time efficiency is enormous for static impurities (almost 40 times faster compared to the conventional Metropolis-Glauber algorithm). For mobile impurities, the time efficiency reduces because of IS moves, which are computationally expensive to implement in the N-Fold way framework. Using the N-Fold way algorithm for mobile impurities, we have studied the nucleation properties starting from $T=1.6$ to $T=0.7$ and observed a qualitative difference in barrier height vs. temperature plot between static and dynamic impurities. This difference arises because of the preferential occupation of impurities at the boundary of clusters at low temperatures. We also observe a discrepancy between direct FFS estimates of the nucleation rate and those computed via US and BDZ theory at low temperatures and for intermediate impurity mobility where the separation of timescales between solute and impurity dynamics is lost.

The presented N-Fold way algorithm provides a useful and powerful tool which could be employed to study nucleation at low temperatures for the 3D lattice-gas models. Also, many experimental conditions consider temperatures well below the critical temperature for such systems. The current methodology could help provide useful context to experimental results and hopefully explore regimes which can be more directly mapped to experimental solution conditions. Moreover, designing the N-Fold way algorithm for a purely diffusive lattice-gas system and studying its nucleation properties can be a potential area for future research. In that case, the PU move can be replaced by a particle swap move between two local or non-local sites.

\section*{Acknowledgements}

DM and DQ were supported by EPSRC grant number EP/R018820/1. Part of the calculations were performed using the Scientific Computing Research Technology Platform at the University of Warwick.

\appendix 

\section{N-Fold way with Kawasaki and particle-update} \label{app:NF}

The algorithm is an extension of the one initially proposed by Bortz et al. \cite{Bortz1975}. The main difference is that, besides PU, IS are managed. The algorithm keeps track separately of both types of moves, considering two different groups of classes, respectively PU-group and IS-group. Each solute/solvent particle ($s_i=\pm 1$) is located in a class of the PU-group depending on its orientation and the sum over the nearest neighbours ($s_j$) of $i$ ($c_i = \sum_j s_j$). The orientation and the coordination number are the only quantities that determine the actual energy difference upon selection of the spin. In particular
\begin{equation}
    \Delta E_i = 2s_i(c_i+h)
    \label{eq:DE_SF}
\end{equation}
In total, 18 classes are needed. Each impurity ($s_i=0$) is located in four classes of the IS-group, one for each of the possible directions of swapping (top, right, bottom, left). The class is selected considering the coordination numbers of the impurity $c_i$ and the swapping neighbouring spin $c_j$. Indeed, the energy difference upon selection of the IS is 
\begin{equation}
    \Delta E_i = s_j(c_j-c_i+1)
    \label{eq:DE_SS}
\end{equation}
In total, 13 classes are needed.

\begin{figure}
    \centering
    \includegraphics[height=3.7cm]{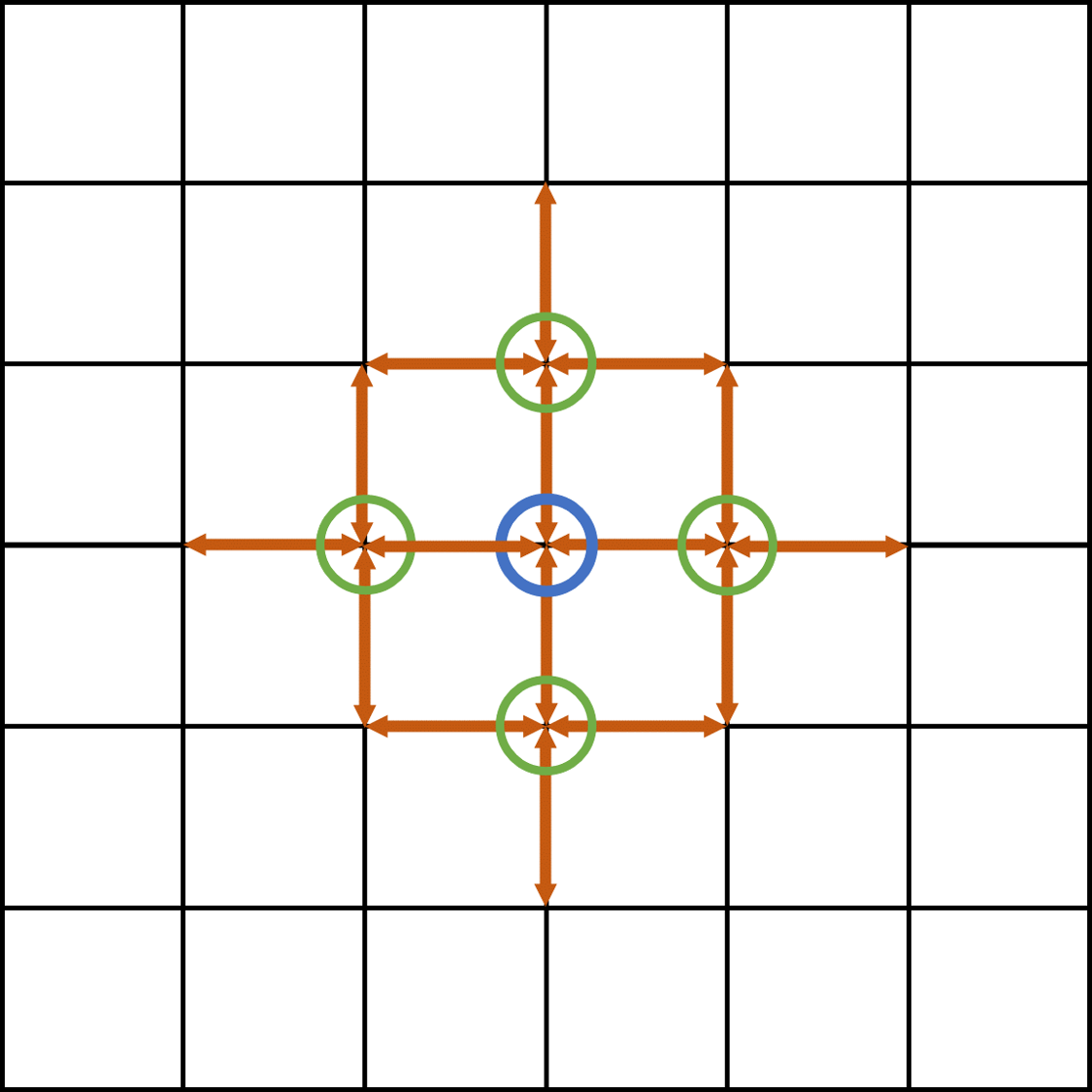} 
    \includegraphics[height=3.7cm]{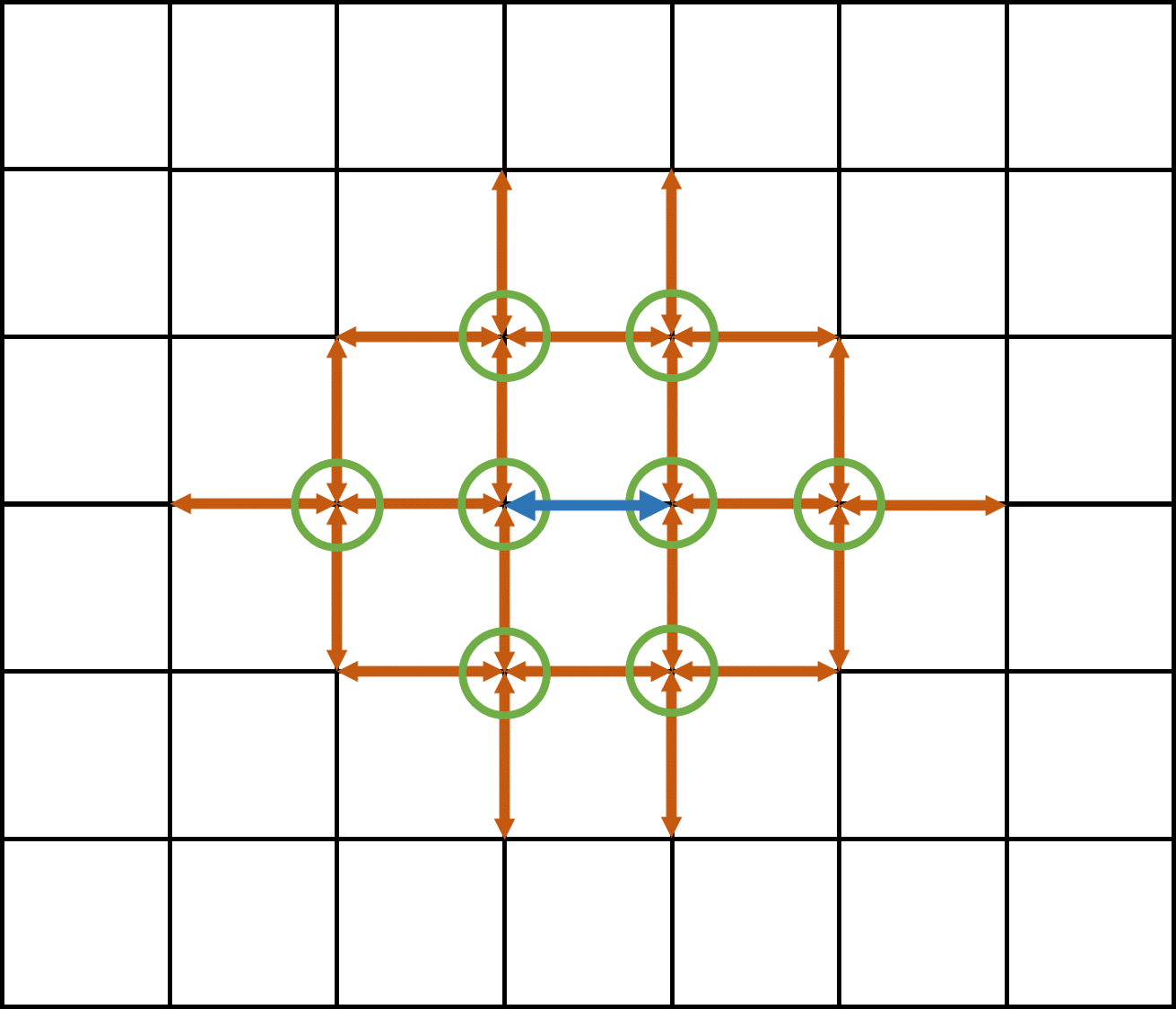} 
    \caption{Left panel: PU move; Right panel: IS move. The blue circle represents the site (sites) that are involved in the selected move. For a PU move, the site variable $s_i$ changes sign, indicating the addition or removal of a particle; for an IS move the two neighbouring sites swap values. Green circles represent sites that need to be checked for PU energy variation consequent to the lattice update step. Orange arrows represent sites that need to be checked for IS energy variation consequent to the lattice update step. }
    \label{fig:Moves}
\end{figure}

After initialisation, the algorithm follows the steps:
\begin{enumerate}
    \item Selection of a move: At each step, a class is chosen, considering both PU and IS, with probability respectively proportional to $\sum_in_i^{PU} w_i^{PU}(\Delta E_i^{PU})$ and $\sum_in_i^{IS} w_i^{IS}(\Delta E_i^{IS})$ where $n_i^{PU/IS}$ represents the total number of elements in the $i$-th class for PU and IS groups and $w_i^{PU/IS}(\Delta E_i^{PU/IS})$ is the Glauber transition rate given by Eqs.~(\ref{eq:Glauber_PU}) and ~(\ref{eq:Glauber_IS}). Note that the selection of an IS move implies the determination of an impurity and a direction of swapping.
    \item Lattice update: The move is applied and the lattice is updated (if a PU is selected, the particle is transmuted; otherwise, the impurity is swapped with the neighbouring site in the chosen direction). Fig.~\ref{fig:Moves} depicts the involved sites in the move with a blue circle for the PU and a blue arrow (involving the swapping sites) for the IS.
    \item Coordination number update: The coordination numbers of each site in contact with the modified sites in the previous step are updated. For an IS move, 8 coordination numbers are modified, for a PU move just 4. The involved sites are depicted in figure \ref{fig:Moves} with a green circle.
    \item Variables update: The class of the sites in the vicinity of the local update are changed according to Eq.~\ref{eq:DE_SF} and ~\ref{eq:DE_SS}. Independently from the type of selected move, both IS-group and PU-group have to be modified. Figure \ref{fig:Moves} depicts the moves (PU in green circles and IS in orange arrows) that should be updated in the surroundings of the selected move.
    \item Time update. Time is increased according to Eq.~(\ref{eq:dt}).
\end{enumerate}

\section{Acceptance Rate Equivalence} \label{app:ARE}
Here we derive the relationship between the time constants $\tau_{IS}$ and $\tau_{PU}$ to establish the equivalence between dynamics of N-Fold way and Metropolis-Glauber algorithms. The dynamics is described by the ratio between the number of applied IS moves and the number of applied PU moves. Therefore, we impose it to be equal for both algorithms. We call $A^{MG}$ the acceptance rate for the Metropolis-Glauber algorithm and $S^{NF}$ the selection rate for the N-Fold way algorithm. We use the subscript PU or IS to indicate the type of move we are referring to. The ratio between the accepted moves and the selected moves can be then written as:
\begin{equation}
    \frac{S_{IS}^{NF}}{S_{PU}^{NF}} = \frac{A_{IS}^{MG}}{A_{PU}^{MG}} 
\end{equation}

For the N-Fold way, we have:
\begin{equation}
    \frac{S_{IS}^{NF}}{S_{PU}^{NF}} = \frac{\sum_i n^{IS}_i \frac{1}{2\tau_{IS}}[1-\tanh{(\beta\Delta E^{IS}_i/2)}]}{\sum_i n^{PU}_i \frac{1}{2\tau_{PU}}[1-\tanh{(\beta\Delta E^{PU}_i/2)}]}
    \label{eq:N}
\end{equation}
where $n^{IS}_i$ and $\Delta E^{IS}_i$ are respectively the number of elements and the energy difference related to the $i$-th class for IS. Analogously for PU.

For the Metropolis-Glauber algorithm, the acceptance rate must be weighted by the probability of generating a specific state. Then, the total rate of accepting  PU move is
\begin{equation}
    A_{PU}^{MG} = (1-\alpha) \sum_j \frac{1}{N_{PU}}\frac{1}{2}[1-\tanh{(\beta\Delta E_j/2)}]
    \label{eq:M_PU}
\end{equation}
where $j$ runs over all the solute and solvent particles in the lattice, $N_{PU} = N(1-f)$ is the total number of particles and $\Delta E_j$ is the energy difference related to the reversal of the $j$-th spin. Here, $(1-\alpha)$ is the probability of selecting all PU states. Analogously, 
\begin{equation}
    A_{IS}^{MG} = \alpha \sum_j \frac{1}{N_{IS}}\frac{1}{2}[1-\tanh{(\beta\Delta E_j/2)}]
    \label{eq:M_SS}
\end{equation}
where $j$ runs over the impurities and $N_{IS} = 4fN$ is the total number of potential IS (accounting also for potential swapping between neighbouring impurities). Here, $\alpha$ is the probability of selecting all IS states. For both Eqs.~(\ref{eq:M_PU}) and~(\ref{eq:M_SS}) we can regroup moves that share the same energy difference, in the N-Fold way. Considering the ratio between these quantities and comparing with Eq.~(\ref{eq:N}), we can simplify the sum over the elements which share the same energetic variation. Then,
\begin{equation}
    \frac{\frac{1}{2\tau_{IS}}}{\frac{1}{2\tau_{PU}}} = \frac{\frac{\alpha}{4fN}}{\frac{(1-\alpha)}{N(1-f)}}
\end{equation}
and hence
\begin{equation}
    \frac{\tau_{PU}}{\tau_{IS}} = \frac{\alpha}{(1-\alpha)}\frac{(1-f)}{4f}
\end{equation}

\section{Cluster Identification Algorithm} \label{app:CIA}
The aim is to follow during the simulation the variation of the size of each geometric cluster made of solute particles. We develop two functions that handle respectively the addition of a particle ($s_i=-1 \to 1$) and the removal of a particle ($s_i=1 \to -1$). A Kawasaki swap requires simply consecutive application of the two functions, in any order, if an impurity swaps with a solute particle. Otherwise, nothing changes. At the beginning of the simulation, we apply the Hoshen-Kopelman algorithm~\cite{Hoshen1976} to list all the clusters and keep track of their size. At the end of the initialisation, three arrays are created.
\begin{enumerate}
    \item A matrix (CLUSTER-NAME): This possesses the same sizes as the lattice. It maps each solute particle to an element of LIST-A.
    \item List of labels (LIST-A): Each cluster has an identification number which is stored in LIST-A. If the identification number is positive, it is interpreted as a pointer to an element of the array LIST-B which gives the cluster's area. If the identification number is negative, then it is interpreted as a pointer towards another element of LIST-A (another cluster) in the position given by the absolute value of the identification number (in Hoshen-Kopelman fashion). The two clusters (pointer and pointed) are merged.
    \item List of sizes (LIST-B): It keeps track of the size of all the clusters in the lattice.
\end{enumerate}
We manage the removal and addition of a solute particle in the following way.
\begin{itemize}
    \item Particle addition: We check the neighbours of the added particles and count the sites which contain a solute particle $m_i$. If $m_i=0$, the solute forms a new cluster and all lists are updated trivially. If $m_i=1$, through CLUSTER-NAME we recover the label of the touched cluster, and we increment its size by one. If $m_i>1$, all independent clusters are merged following the Hoshen-Kopelman algorithm.
    \item Particle removal: As earlier, we check the neighbours of the subtracted particle $s_i$ and count the number of nearest neighbour sites that are solute particles $m_i$. As for the addition process, the cases $m_i=0$ or 1 are handled trivially. More complex is the case $m_i>1$, since we must discriminate when the subtraction of a solute particle breaks the cluster into two (or more) independent ones (demerging of the cluster). The only way to exclude the demerging of clusters is to reconstruct simultaneously, solute particle after solute particle, the shape of each sub-cluster and to look for other points of contact between them. Once a point of contact is found, it means that the touching clusters belong to the same cluster and are treated as so. The reconstruction process proceeds until all clusters are merged or until no points of contact are found (and no more solute particles can be added to the remaining clusters). The first condition can be met early in the reconstruction process, as it happens when a solute particle is subtracted from the centre of a cluster. The second process, instead, usually requires more steps: to exclude the merging all particles must be considered and their neighbours must be checked. In more detail, the algorithm proceeds in the following way. Each neighbouring site containing a particle $s_j=1$  is written as the first element in a list, for a total of $m_i$ lists. We pick the first site of the first list and we add to that list all the neighbouring sites which have a solute particle. Then we go to the second list and do the same. This is done until all $m_i$ lists have been considered. Then we start back from the first list considering the second element. In general, two counters are considered: one for the list and one for the element in the list. We scan through all lists in the same position and then we increment the element counter starting back from the first list. The procedure is iterated until no more particles can be added to a growing cluster or two of them touch (a site appears in more than one list). In the former case, the isolated cluster acquires a new label in LIST-A and its size is recorded in LIST-B. In the latter case, the isolated clusters that touch are merged into one and the growth continues. When only one growing cluster is remaining the demerging algorithm stops. The remaining cluster's area is then the difference between the initial cluster's area and the areas of the merged and isolated clusters minus 1. LIST-A and LIST-B are updated accordingly.
\end{itemize}

\bibliography{MyBibliography}
\bibliographystyle{elsarticle-num}

\end{document}